# THE CASE FOR CONTROLS: IDENTIFYING OUTBREAK RISK FACTORS THROUGH CASE-CONTROL COMPARISONS

**Authors**

Nina H. Fefferman[1-4,*], Michael J. Blum[1,*], Lydia Bourouiba[5], Nathaniel L. Gibson[1], Qiang He[6,7], Debra L. Miller[4], Monica Papeş[1], Dana K. Pasquale[8,9], Connor Verheyen[5], Sadie J. Ryan[10-12]

**Affiliations**

[1] University of Tennessee, Department of Ecology and Evolutionary Biology, Dabney Hall, 1416 Circle Dr, Knoxville, TN, 37996, USA

[2] University of Tennessee, National Institute for Mathematical and Biological Synthesis, 1122 Volunteer Blvd, Suite 106, Knoxville, TN, 37996, USA

[3] University of Tennessee, Department of Mathematics, 1403 Circle Dr, Knoxville, TN, 37996, USA

[4] University of Tennessee, One Health Initiative, 2621 Morgan Circle, Knoxville, TN, 37996, USA

[5] The Fluid Dynamics of Disease Transmission Laboratory, Fluids and Health Network, Massachusetts Institute of Technology, 77 Massachusetts Ave, Cambridge, MA, 02139, USA

[6] The University of Tennessee, Department of Civil and Environmental Engineering, 325 John D. Tickle Engineering Building, 851 Neyland Drive, Knoxville, TN, 37996, USA

[7] The University of Tennessee, Institute for a Secure and Sustainable Environment, 600 Henley Street, Suite 311, Knoxville, TN, 37996, USA

[8] Duke University School of Medicine, Duke Department of Population Health Sciences, 215 Morris St, Durham, NC, 27701, USA

[9] Duke University, Duke Network Analysis Center, Box 90989, Durham, NC, 27708, USA

[10] University of Florida, Department of Geography, Quantitative Disease Ecology and Conservation (QDEC) Lab, 3141 Turlington Hall, 330 Newell Dr, Gainesville, FL, 32611, USA

[11] University of Florida, Emerging Pathogens Institute, 2055 Mowry Rd, Gainesville, FL, 32610, USA

[12] University of KwaZulu-Natal, College of Life Sciences, Room 03-004, New Biology Building, Westville Campus, Westville, 3629, Durban, South Africa

*: Co-First authors

**Corresponding Author**
Nina Fefferman
University of Tennessee
Department of Ecology and Evolutionary Biology
Knoxville, Tennessee 37996-3140
Email: nina.h.fefferman@gmail.com

**Keywords:** Evidence-based Epidemiology, Outbreak investigation, Case-control epidemiology, epidemiological hierarchy of evidence

**SUMMARY**
Investigations of infectious disease outbreaks often focus on identifying place- and context-dependent factors responsible for emergence and spread, resulting in phenomenological narratives ill-suited to developing generalizable predictive and preventive measures. We contend that case-control hypothesis testing is a more powerful framework for epidemiological investigation. The approach, widely used in medical research, involves identifying counterfactuals, with case-control comparisons drawn to test hypotheses about the conditions that manifest outbreaks. Here we outline the merits of applying a case-control framework as epidemiological study design. We first describe a framework for iterative multidisciplinary interrogation to discover minimally sufficient sets of factors that can lead to disease outbreaks. We then lay out how case-control comparisons can respectively center on pathogen(s), factor(s), or landscape(s) with vignettes focusing on pathogen transmission. Finally, we consider how adopting case-control approaches can promote evidence-based decision making for responding to and preventing outbreaks.

**INTRODUCTION**
Understanding disease emergence and spread is a complex challenge that transcends current modes of investigation. Outbreaks are typically investigated through expository approaches that aim to identify place- and context-dependent causative factors. As the focus is on understanding the basis of a specific outbreak, the resulting phenomenological 'just so' narratives are rarely suitable for forecasting risk or developing generalizable predictive and preventative measures. While it is widely understood that factors rarely act in isolation, conventional approaches can fall short when disease dynamics arise from synergies among multiple factors, warranting study from multiple disciplinary perspectives. Contributing factors can be firmly rooted in virology, immunology, and pharmacology [1], or may describe ecological and environmental influences [2], while others might relate to social and engineering domains examining human actions, behaviors, and interventions [3,4]. Phenomenological investigations of particular outbreaks also can fall short of capturing the contingent nature of multifactorial dynamics, where the importance of one factor can be contingent on a "perfect storm" of other concurrent factors. Here we contend that adopting a multidisciplinary case-control hypothesis testing framework can substantively advance understanding of disease emergence and spread.

Great strides have been made in developing new approaches to investigate outbreaks. For example, there has been notable progress in capturing complex dynamics in landscape-based epidemiological modeling of outbreaks. Whereas initial models considered a particular outbreak scenario of interest and hypothesized the contributions of observable sets of features [5–9], more recent efforts examined multiple outbreaks [10–12] and have utilized techniques such as fore- and hindcasting to test predictive models relying on either statistical patterns [13–15] or inferred causal dynamics, as discussed in [16]. While these approaches have provided valuable insights, it is unclear whether they will provide a robust basis for studying and forecasting complex outbreaks.

We propose that a key step in advancing understanding of disease outbreaks is the development and adoption of a case-control analytical toolkit [17]. Borrowing from other investigative traditions, case-control epidemiological analysis can serve as a platform for iterative rounds of hypothesis generation and testing. By analyzing combinations of (perhaps conditionally dependent) factors from across disciplines, the approach allows for the identification of minimally sufficient sets of factors (Box 1) that tip emergent infections over into outbreaks.

> **Box 1**
> **MINIMALLY SUFFICIENT SETS OF POTENTIAL CAUSAL FACTORS**
> Consider any case in which the objective is to study what leads to a positive versus negative outcome (i.e., an outbreak versus no epidemic spread). Suppose also that we can describe a set, $S$, of elements 1 through $N$ of possible causal factors that we hypothesize might together determine the outcome. A sufficient set, $M_j$, is any subset $S$ that, taken together, leads to a positive outcome. Mathematically, if $S = \{x_1, \ldots, x_N\}$ where each element $x_i = \{0 \ if \ absent/false \ 1 \ if \ present/true \}$, then a sufficient set is a subset $M \subseteq S$ such that if $\forall y_i \in M_j, y_i = 1$ then the studied outcome is guaranteed to be positive. Further, M is minimally sufficient if for any $y_i \in M_j, y_i = 0$, then the studied outcome is guaranteed to be negative (i.e., a subset is minimally sufficient if the exclusion of any one of the elements changes the outcome from positive to negative). Note that one minimally sufficient set cannot be a subset of another, but there is otherwise no restriction on overlap between minimally sufficient sets, nor on the number of elements contained in the minimally sufficient set. For example, if $N = 10$, then we could have $M_1 = \{x_1, x_2, x_3\}$, $M_2 = \{x_1, x_2, x_4\}$, $M_3 = \{x_1, x_2, x_5, x_6, x_7\}$, $M_4 = \{x_8\}$, and $M_5 = \{x_9, x_{10}\}$ all as minimally sufficient sets. If those are all of the minimally sufficient sets contained within S, then we have a maximal set of minimally sufficient subsets and, therefore, a full description of all the ways in which we could have a positive outcome.

Modern medical research serves as a valuable precedent, illustrating the merits of adopting case-control comparisons to investigate multifactorial dynamics. Medical research shifted from investigating pathology in the human body as a set of independent tests of hypotheses based on expert opinion and careful observation towards employing a framework for evidence-based inquiry that forces inclusion of otherwise unanticipated drivers and that provides a pathway for synthesis within a hierarchy of evidence [18]. Within that hierarchy are tools designed specifically to move scientific understanding from enigmatic, heterogeneous, multifactorial data to well-supported, evidence-based understanding of the drivers of pathology. A basic tool of evidence-based medicine is the concept of the Case-Control Study [17]: a retrospective analysis in which individuals who exhibit an observed outcome (cases) are "paired" in study design with individuals who do not exhibit the observed outcome (controls), with analysis then performed to identify which exposures (i.e., potential driving factors) they share and which are distinct only among the cases. This allows for the calculation of measures such as odds ratios [19] to estimate the likely impact of suites of features on health outcomes.

Analogous to the way medicine concerns pathologies of the body, epidemiology considers outbreaks as pathologies of a population. Unlike physiology, however, a multitude of potential types of controls may be matched to outbreak cases for exploration and analysis. To isolate and understand epidemiological risk and the factors that drive (re)emergence of pathogens of concern, we propose three main categories of Case-Control formulation that can serve as a general foundation for evidence-based epidemiological analysis of outbreak risk: (1) Factors; (2) Pathogens; and (3) Landscapes. In each of these categories, the observed outcome that exists for Cases and that does not for Controls is a circulating outbreak, but what constitutes a well-matched Control is partitioned into a narrower equivalence class.

**CASE CONTROL: FACTORS**
Evidence-based epidemiological analysis of outbreak risk focuses on factors that may limit or foster a disease outbreak (Box 2). Abiotic factors such as temperature, humidity, and precipitation can define the distribution and abundance of a pathogen. Contextual or geographic factors capturing socioeconomic conditions, aspects of the built environment, or human demography and behavior (e.g., lockdowns) may also influence pathogen transmission risk.

**Temperature**. Dengue fever virus is a flavivirus transmitted by *Aedes aegypti* and *Aedes albopictus* mosquitoes. Originally of sylvatic origin, the virus is now characteristically found in the human environment, with seasonal to perennial circulation among container-breeding mosquitoes and humans [20]. When symptomatic in humans, infection causes high fever, joint aches, and oftentimes has a characteristic macular rash [21]. Typically thought of as a tropical disease, dengue has been (re)emerging in novel extra-tropical locations as climate change has been shifting temperature bounds on transmission suitability [22]. Accordingly, temperature could limit transmission to locations where the environment is sufficient to support the vector, *Aedes spp.* mosquitoes. A case-control pairing could be established based on presence or absence of suitable temperature for the *Aedes* vector(s). By extension, a related factor in prevention of outbreaks, such as knowledge of dengue risk, could be assessed by exploring it in a city well within the temperature bounds of suitability, versus a city whose average temperatures never or rarely exceed the minimum bounds for transmission suitability.

**Sociopolitical and economic (in)stability.** In the former Soviet Union, childhood vaccination campaigns for diphtheria were challenged when adults experienced waning immunity [23]. Diphtheria, caused by the bacterium *Corynebacterium diphtheriae*, can form an obstructive biofilm that hinders breathing and swallowing. It can also produce a blood-borne toxin that can cause fatal heart and nerve damage. Disruption of childhood vaccination protocols during the fall of Soviet Union led to widespread and unanticipated outbreaks of diphtheria among adults due to waning immunity [23]. The duration of childhood immunity also degraded more rapidly and for a greater percentage of vaccination recipients than anticipated [24]. This effectively served as a natural case-control experiment, contrasting the same population against itself in a scenario of demographically dependent herd immunity. When outbreaks were sufficiently prevented among children, the entire population was largely protected against widespread transmission, but as the demography of vaccine protection shifted, so did outbreak dynamics. Socioeconomic instability was also an obvious contributing factor, limiting access to transmission-blocking care. Disruptions of an intervention aimed at population-level protection (e.g., vaccination, water treatment, food aid) as a function of sociopolitical and economic instability are not unusual: treating the 'before and after' in a time-series of health outcomes as case-control comparisons can provide evidentiary support for long-term investment in population-level health and nutrition programs.

**CASE CONTROL: PATHOGENS**
Evidence-based epidemiological analysis of outbreak risk might focus explicitly on a pathogen of interest, with case-control comparisons examining biological conditions that may limit or foster a disease outbreak. Here, the focus might be on the nature of distributions, relative abundance, or ecological interactions that shape pathogen transmission or exposure risk.

*Trypanosoma cruzi*. Chagas disease is an anthropozoonosis caused by *Trypanosoma cruzi*, a protozoan parasite transmitted to humans and other mammals via triatomine 'kissing' bugs. Chagas is a global health threat, with the current majority of cases primarily concentrated in Central and South America. It is estimated that eight million individuals are currently infected, and upwards of 100 million people live in areas of high infection risk [25]. While Chagas has historically been considered a disease of rural communities, there is growing evidence that urban populations are at risk of *T. cruzi* infection. [26,27]. Evidence of *T. cruzi* in cities implies the presence of triatomine vectors that sustain infection in peridomestic reservoir hosts (e.g., rodents, racoons, dogs), but triatomines have been sparingly observed in urban landscapes. A case-control approach could test the hypothesis of transient infection of *T. cruzi* in cities, with consideration given to conditions that can sustain infection in hosts where triatomines are rare or absent. This could reveal novel mechanisms contributing to infection (e.g., vertical transmission) and offer new insights about the eco-epidemiology of one of the world's most burdensome diseases.

**Hantaviruses.** Hantaviruses that cause human disease persist enzootically in some species of rodents, maintained via horizontal transmission among reservoir species [28,29]. The virus is excreted in urine and feces, with human infection occurring most commonly via inhalation of aerosolized matter containing the virus inside dwellings [29] and, as treatment comprises only supportive care, can lead to death [29]. Each of the hantaviruses known to infect humans correspond to a single rodent host genus [30], but viral assortment occurs, potentially generating novel strains [30]. Factors associated with human infection include environmental conditions (e.g. rainfall, temperature), reservoir population density affecting viral prevalence, and variation in human-reservoir interactions across different landscapes [28]. A case-control approach for hantavirus could therefore test for the presence (i.e., external to and inside of human dwellings) and population size of a particular rodent reservoir. This approach could reveal human-reservoir interaction tipping points as a function of the number and nature (duration, frequency, e.g.) of exposures leading to human infection.

**CASE CONTROL: LANDSCAPES**
Finally, evidence-based epidemiological analysis of outbreak risk might consider landscape-level phenomena that shape disease emergence and spread. As illustrated below, consideration might be given to landscape functionality and features such as degree and type of habitat connectivity.

**Public water systems.** *Cryptosporidium* is an apicomplexan parasite that causes the diarrheal disease cryptosporidiosis. Humans may be infected as a result of exposure to fecal matter shed by animals carrying certain species of *Cryptosporidium* [31]. Waterborne *Cryptosporidium* is very resistant to chlorine, the most common disinfectant in the United States [32]. Consequently, outbreaks of cryptosporidiosis can result from exposure to *Cryptosporidium* in drinking water, potentially impacting large populations served by public water systems in the United States. The landscape can be a critical determinant of the transport of fecal matter from infected animals, such as livestock in concentrated animal feeding operations (i.e., CAFOs), to the sources of public water systems. Landscape features may include proximity of the source water to infected animals, weather and land cover conditions that are conducive to stormwater runoff (i.e., connectivity and transport), and inadequate source water management practices (i.e., containment or removal). Accordingly, a case-control analysis could compare populations served by public water systems with distinctive landscape features to identify causes of cryptosporidiosis outbreaks.

**Social-ecological mosaics.** Cities can be described as social-ecological landscapes, wherein there is an admixture of natural and built spaces adjoining or overlapping with one another. Within social-ecological landscapes, habitat mosaics can reflect differences in land use, habitation, disturbance (e.g., natural disasters), and economic development, sometimes reflecting disparities driven by discriminatory public policy [33–35]. Spatial patterns in outbreak preparedness, perhaps driven by regional beliefs, can shape exposure risks [36]. Phenomena like property abandonment and persistent vacancy can create social-ecological mosaics that influence risks of exposure to peridomestic animals that host pathogens of concern [35]. It has been demonstrated, for example, that the diversity, distribution, and prevalence of rodents and associated pathogens can reflect mosaics of abandonment in neighborhoods within a city [33,34]. Notably, these patterns can directly tie exposure risk to spatial socio-economic landscapes, creating hotspots of risk in already historically underserved areas. Case-control analysis of rodent demography in different neighborhoods could reveal how abandonment shapes exposure risk among neighborhoods within cities [33,34] and between cities (i.e., exhibiting distinct patterns of abandonment) to consider how social-ecological forces manifest risk across distinct geographies.

**DISCUSSION**
While many factors driving outbreak risk are well-studied, more work is warranted to better understand what allows an infectious disease to emerge and then tip from endemic to epidemic to pandemic. We contend that greater knowledge and predictive capacity can be achieved by employing a case-control framework for inquiry that forces consideration of multidisciplinary drivers of disease outbreaks, providing a pathway towards synthesis across a hierarchy of evidence (Box 2).

**Box 2**

**CASE CONTROL APPROACH IN PANDEMIC RESPONSE – SEASONALITY, RELATIVE HUMIDITY AND COVID-19.**

The severity of infectious respiratory diseases, such as Influenza or COVID-19, is often suggested to be seasonal, though the evidence supporting a relationship between weather and respiratory infectious disease outbreaks remains controversial [38,39]. Any causal relationship between weather and respiratory infectious diseases is expected to be complex and multifaceted, involving rich biophysical and fluid physics processes as well as host and pathogen physiology modulated by environmental factors [40]. Research on this topic has thus far primarily focused on geographically localized datasets and outdoor conditions. Case-control analysis could be a more suitable and even necessary approach to shed light on the nature of respiratory disease dynamics. For example, the pre-vaccine COVID-19 pandemic stage covered a wide range of climates and is thus suitable for contrasting regional conditions. Verheyen and Bourouiba conducted a large-scale retrospective analysis blending both case-control and cohort methodologies across a comprehensive global dataset comprising COVID-19 statistics, outdoor meteorological variables, and experimentally-validated indoor relative humidity (RH) estimates. The study revealed a robust relationship between regional indoor RH and COVID-19 outbreak severity. The study also revealed that COVID-19 outcomes during the early stages of the pandemic (i.e., pre-vaccination) were consistently less severe at intermediate indoor RH between 40% and 60% and more severe at extreme indoor RH outside of this range. The association remained robust even after controlling for outdoor weather conditions, government response, and the statistical methodologies and data processing approaches used. Caution is essential, however, when conducting global case-control analyses. Inter-country variability in outbreak magnitudes and reporting conditions can lead to noisy, error-prone datasets, indicating a need for normalization to ensure validity of comparisons. Thus, the very notion of "robustness" ought to include extensive sensitivity analyses (e.g., [41]), covering differing treatments and assumptions for both the underlying datasets and the methodology. Emerging associations should persist even when many of the assumptions or methods of analysis are perturbed.

**Figure Caption:**

**FIGURE 1 IN BOX 2: (a)** The aggregated global dataset was stratified in color-coded regions (map) to investigate the effects of low indoor relative humidity (RH) in contrast to moderate and high RH. **(b)** The analysis reveals a U-shaped curve of association between indoor RH and new COVID-19 deaths with a sweet spot range of indoor RH associated with reduced death. Case-control and cohort methods were used to further refine the potential relationship between COVID-19 outbreaks and indoor climate with odds ratios supporting further association between intermediate indoor RH and better global outcomes like fewer deaths, a smaller increase of deaths, or a smaller percentage change of death **(c-e).** The consistent findings of Verheyen and Bourouiba [41] may be a function of the high-level nature of the analysis, where data aggregation masks lower-level variations while enabling the extraction of the general most robust patterns in the data [42]. Extensive sensitivity analyses are warranted due to the inherent heterogeneity of epidemic datasets across regions and timeframes. In Verheyen and Bourouiba [41], comprehensive robustness tests were performed, including assessments of changes in outdoor conditions and government interventions as well as direct experimental verification of the method used to convert outdoor to indoor humidity.

Case-control formulations allow for more rigorous and explicit consideration of complex interactions among sets of factors (Box 1) that may not yet be fully identified or understood; generate hypotheses for further investigation; and bridge gaps among disciplines in ways that are otherwise difficult to synthesize. We illustrate the versatility and merits of a case-control approach to promote evidence-based decision-making. We further argue the merits of isolating minimally sufficient sets of factors. This allows us to move away from a more traditional focus on identifying and reducing one large single-source risk, since that may inadvertently create equal, if not greater, risks from a diffuse suite of other, now slightly altered contributory factors [37]. While many of the factors driving outbreak risk have been well studied, much more work is warranted to better understand the myriad of ways in which an infectious disease can tip from endemic to epidemic to pandemic. Employing a more robust analytical framework and investigative toolkit can only aid in ongoing efforts to progress beyond phenomenological interrogation and to establish a consistent basis for developing generalizable preventable measures that respond to dynamic, multifactorial risks.

## ACKNOWLEDGMENTS
We thank our collaborating members of the Predicting Emergence in Multidisciplinary Pandemic Tipping-points (PREEMPT) Institute for useful discussion and feedback about this effort.

## DATA SHARING
No data were collected for this study (i.e., theoretical, review, opinion, editorial papers).

## DECLARATION OF INTERESTS
No authors who contributed to this manuscript have conflicts of interest.

## FUNDING
This material is based upon work supported by the National Science Foundation under Grant CCF #2200140. The National Science Foundation was not involved in this study.

## AUTHORS' CONTRIBUTIONS
NHF, MJB, LB, and SJR each contributed to the funding acquisition. NHF, MJB, LB, NLG, QH, DLM, MP, DKP, and SJR all contributed to the conceptualization, and to writing the original draft, and then reviewed and edited the manuscript. LB and CV each contributed to the visualization.

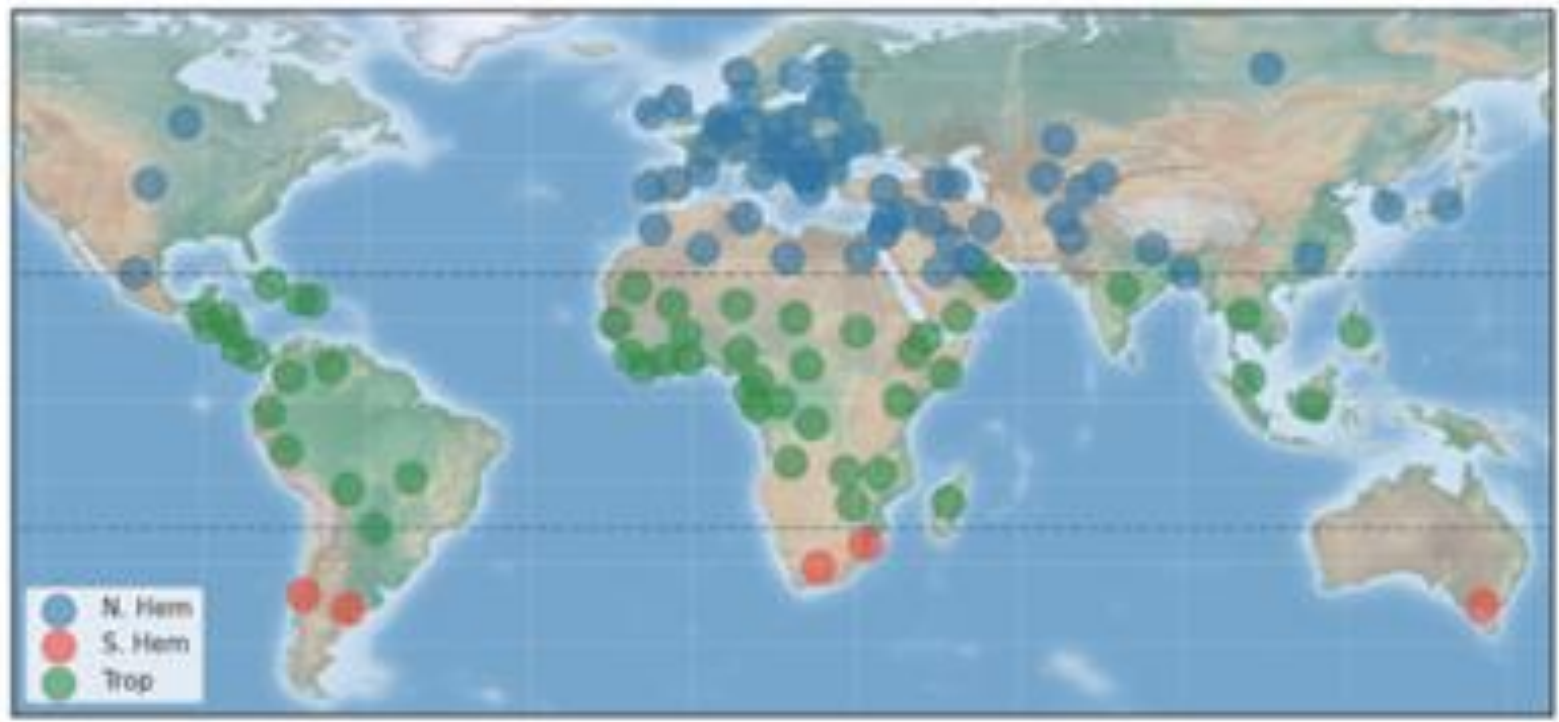
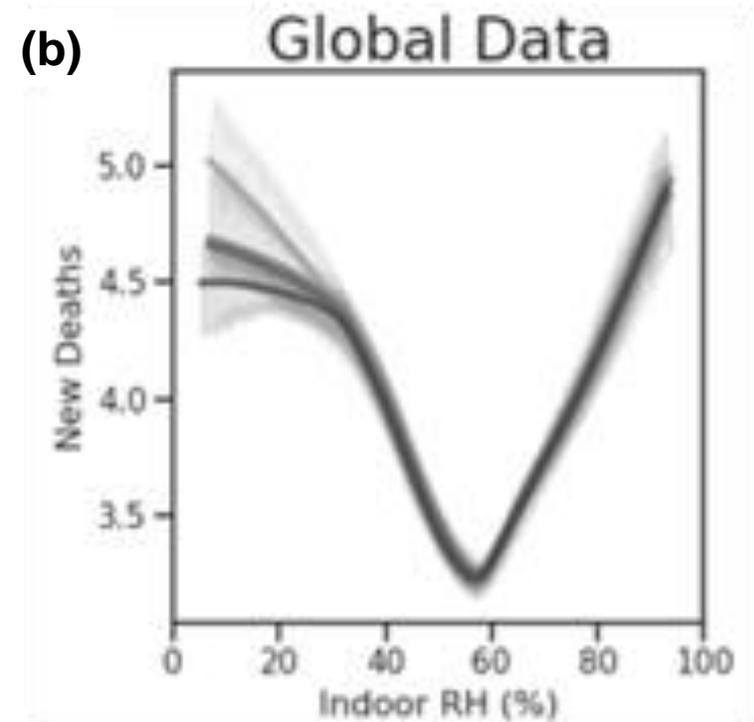
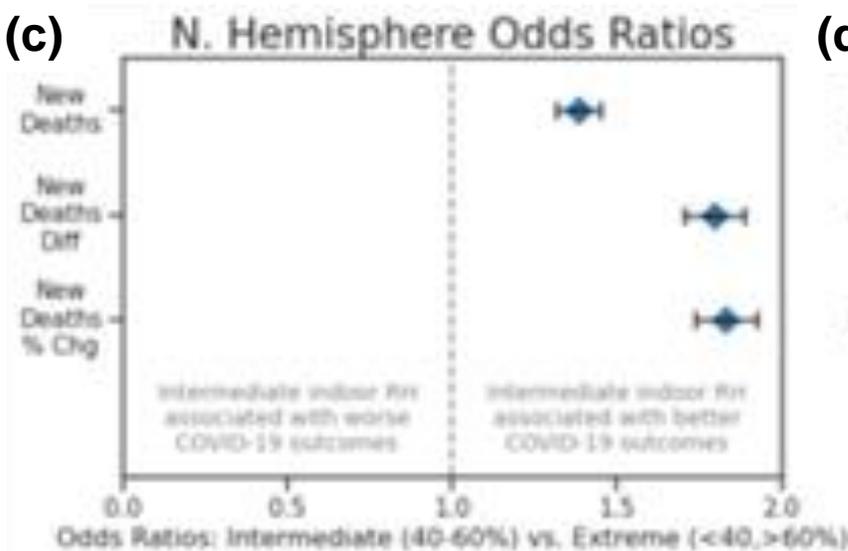
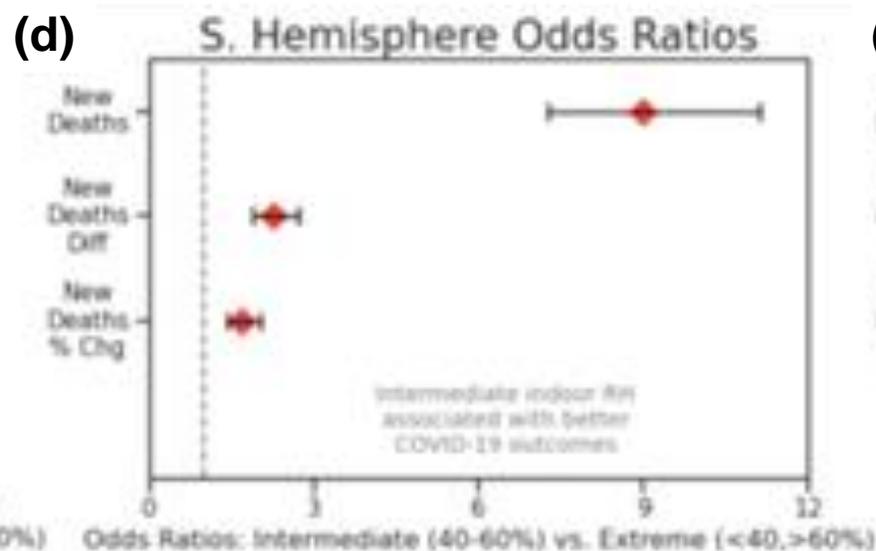
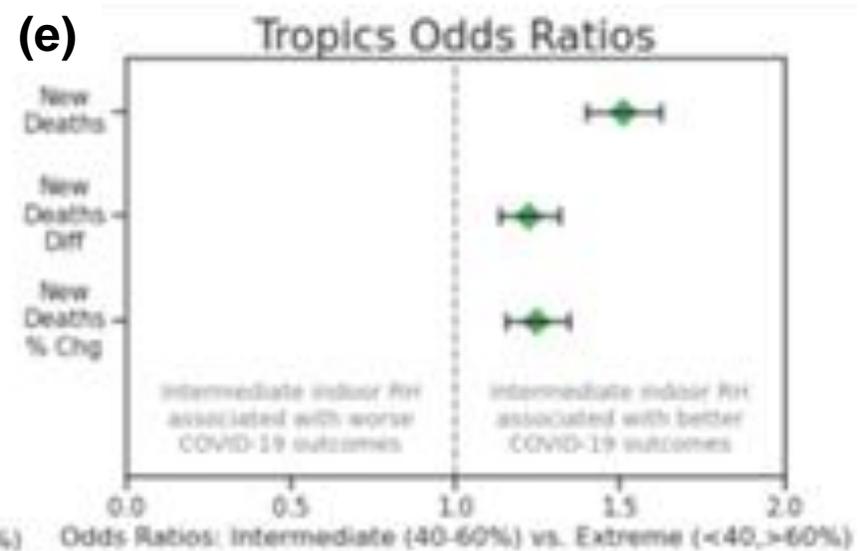

Figure 1